\documentstyle[12pt,axodraw]{article}
\def\pa{\parallel}
\def\pe{\bot}

\def\bea{\begin{eqnarray}}
\def\eea{\end{eqnarray}}
\def\be{\begin{equation}}
\def\ee{\end{equation}}
 \let\b=\beta   \let\d=\delta  \let\e=\varepsilon
        
\let\m=\mu   \let\n=\nu   \let\o=\omega    
   \let\t=\tau   
  \let\z=\zeta
\let\D=\Delta   \let\G=\Gamma

\title{\bf Renormalized Field Theory of Infinitely 
Driven Lattice Gases}

\author{F. de los Santos $^1$ and Miguel A. Mu\~noz $^{1,2}$
 \\ $^1$
 Institute {\em Carlos I} for Theoretical and Computational Physics\\ 
and Departamento de Electromagnetismo y F\'\i sica de la Materia\\
 18071 Granada, Spain.\\
 $^{2}$ The Abdus Salam International
 Centre for Theoretical Physics\\
P.O. Box 586, 34100 Trieste, Italy}
\begin{document}
\maketitle
\begin{abstract}
We use field theoretic renormalization group methods to study
the critical behavior of a recently proposed Langevin equation
for driven lattice gases under
infinitely fast drive.
 We perform 
an expansion around the upper critical
dimension, $d_c=4$, and obtain the critical exponents 
to one loop order. The main features of the 
two loop calculation are then outlined.  
The renormalized theory is shown to exhibit a behavior 
different
from the standard field theory for the DLG, {\em i.e.\/}
it is not mean field like.
\end{abstract}

Since it was first introduced by Katz {\em et al.\/} \cite{katz},
the {\em driven lattice gas\/} model (DLG hereafter) has attracted 
considerable interest \cite{zia,marro}. 
Being one of the simplest archetypes
of non-equilibrium model its study may contribute to
pave the way for an understanding of out of equilibrium systems.
The DLG consists of a periodic regular lattice on which 
nearest-neighbor particle-hole exchanges are performed. The hopping rate
is determined by the energetics
 of the Ising Hamiltonian $H$, the coupling to
a thermal bath at temperature $T$, 
and an external uniform driving field
$\bf E$ pointing along a specific lattice axis. 
In particular,
the hoping rate depends on 
$[(\D H+ \ell E)/T]$, where $\D H$ is the energy variation 
which would be caused by 
the tried configuration change, $E=|{\bf E}|$, and
$\ell=1$ (-1) for jumps along (against) $\bf E$ and 0
otherwise (see \cite{zia,marro} for a detailed description).
The DLG exhibits a 
continuous phase transition 
from a disordered state at high $T$ to a stripe like ordered state at 
sufficiently low $T$ \cite{zia,marro}.  The nature and properties of this
transition have been largely studied in recent years. 
A new general Langevin equation has been proposed,
capturing the physics of the DLG at the critical point \cite{pre}.
 This Langevin equation
predicts different critical
behavior for the cases $0<E<\infty$ and
 $E=\infty$ (in which particles cannot jump against the field) 
respectively \cite{pre,jsp}. 
That is, the point $E=\infty$ behaves as
a sort of {\it tricritical point} in the parameter space.
For finite values of the driving field $E$ the 
Langevin equation previously proposed by 
Janssen and Schmittmann is recovered \cite{janssen}.
It is the purpose of this paper to investigate the critical behavior 
of the DLG for $E=\infty$ in order to determine
explicitly the differences with the $0<E<\infty$ case.
The new
Langevin equation reads \cite{pre,jsp}
\bea
\label{new}
\partial_t \phi ={e_0 \over 2}
\Big[ -\D_\pa \D_\pe \phi-\D_\pe^2 \phi+
\t \D_\pe \phi+ {g \over 3!} \D_\pe \phi^3 \Big] \nonumber \\
+\sqrt{e_0} \ \nabla_\pe \cdot \mbox{\boldmath $\xi$}_\pe
+\sqrt{e_0 \over 2} \ \nabla_\pa \xi_\pa,
\eea
where $\nabla_\pa $ ($\nabla_\pe$) is 
 the gradient operator in 
the direction parallel (perpendicular) to the electric field, and
the noise satisfies
\bea
\langle \mbox{\boldmath $\xi$}({\bf x},t) \rangle &=&0, \nonumber \\
\langle \mbox{\boldmath $\nabla$}
\cdot \mbox{\boldmath $\xi$}({\bf x},t)
\cdot \mbox{\boldmath $\nabla$}' \cdot
\mbox{\boldmath $\xi$}({\bf x}',t') \rangle
&=&-\mbox{\boldmath $\nabla$}^2 \d({\bf x}-{\bf x}') \d(t-t').
\eea
A very similar equation has been proposed to describe the Freedericksz transition in nematic liquids, 
and general asymmetric two-dimensional pattern formation \cite{nematic}.

In order to renormalize this equation,
 following standard methods \cite{bausch},
let us introduce a Martin-Siggia-Rose
response field $\tilde \phi$ and recast 
Eq. (\ref{new}) as a dynamical functional \cite{GF}, the
associated action of which is
\bea
{\cal L}(\tilde \phi, \phi)= \int d^dx dt \Bigg\{
\tilde \phi \Big[ \partial_t -{e_0 \over 2} (-\D_\pa \D_\pe
-\D_\pe^2 +\t \D_\pe) \Big] \phi
-{e_0 \over 2} {g \over 3!} \tilde \phi \D_\pe \phi^3 \nonumber \\
-{e_0 \over 2} \tilde \phi \Big (\nabla_\pe^2 + {1 \over 2}
\nabla_\pa^2 \Big)
\tilde \phi \Bigg\}.
\label{funcional}
\eea

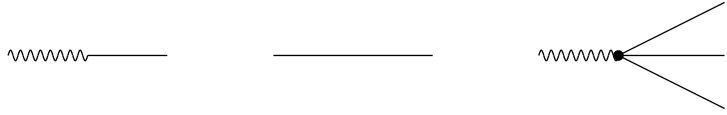
\begin{figure}
\begin{center} \begin{picture}(270,65)(0,0)
\Photon(0,50)(30,50){2}{8}
\Line(30,50)(60,50)
\Line(100,50)(160,50)
%
\Photon(200,50)(230,50){2}{8}
\Line(230,50)(270,50)
\Line(230,50)(270,70)
\Line(230,50)(270,30)
\Vertex(230,50){2}
\end{picture}
\end{center}

\caption{Elements of perturbation theory: the response and correlation 
propagators and the four-point vertex. $\tilde \phi$-legs are indicated by 
a wiggle.}
\end{figure}
 
The free propagators are 
\bea
G_{02}^0({\bf k},\o) &=& {-e_0 ( k_\pe^2+{1 \over 2}k_\pa^2) \over
\o^2+({e_0 \over 2})^2 k_\pe^4(k^2 + \t)^2},
\nonumber \\
G_{11}^0({\bf k},\o) &=& {1 \over i\o+{e_0 \over 2} k_\pe^2(k^2 + \t)},
\eea
and the vertex is: 
$- e_0 g /12 \ k_\pe^2$. 
These elements can be represented diagrammatically as in 
Figure 1 (wavy legs symbolize response fields; straight 
lines stand for density fields).

In order to renormalize the theory, one has to look for the primitive
divergences in a perturbation expansion.
If $\G_{\tilde n n}$ denotes a one-particle irreducible vertex
function with $\tilde n$ external $\tilde \phi$-legs and $n$
external $\phi$-legs, only $\G_{11}$ and $\G_{13}$ are found
to possess primitive divergences. The Feynman diagrams contributing
to these vertex functions are topologically identical to model B 
graphs \cite{bausch}.
 However, the bare correlation and response propagators that
follow from  Eq. (\ref{funcional}) are anisotropic, in contrast to their
counterparts in model B \cite{bausch}. 

\begin{figure}[t]
\begin{center} \begin{picture}(300,100)(0,0)
\Photon(10,20)(50,20){1}{8.5}
\Line(50,20)(90,20)
\CArc(50,50)(30,0,360)
\Vertex(50,20){2}
\PhotonArc(240,50)(30,0,90){1}{8.5}
\CArc(240,50)(30,90,360)
\Photon(170,80)(210,50){1}{8.5}
\Line(170,20)(210,50)
\Line(270,50)(310,80)
\Line(270,50)(310,20)
\Vertex(210,50){2}
\Vertex(270,50){2}
\end{picture}
\end{center}

\caption{One loop diagrams contributing to
$\G_{11}$ (left) and $\G_{13}$ (right).}
\label{diagdiv}
\end{figure}
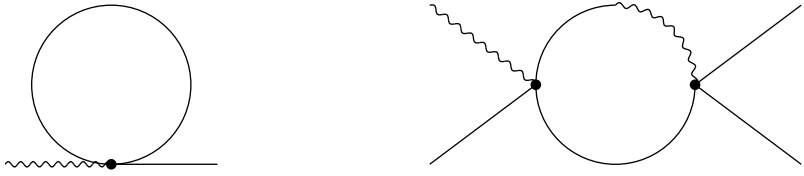

To one loop in $\e=4-d$, the ultraviolet divergences in $\G_{11}$ and 
$\G_{13}$ lead to the renormalization of $\t$ and $u$, the latter
being the dimensionless coupling constant $u \equiv A_{\e} \t^{-\e/2} g$.
$A_\e$ is a numerical factor to be defined below.
We define renormalized parameters $\t_R$ and $u_R$ by
$\t_R = Z_\t \t$ and 
$u_R = Z_u u $.
Given that the leftmost diagram in Figure 2 does not depend
on external moments or frequencies, the derivatives of $\G_{11}$ with respect
to them vanish, and no extra (field) renormalizations are 
therefore required.
The $Z$ factors are determined by the following normalization conditions
\bea
\partial_{k_\bot^2} \G_{11}^R \vert_{NP} &=& {e_0 \over 2} \t_R, \nonumber \\
\partial_{k_\bot^2} \G_{13}^R \vert_{NP} &=& {e_0 \over 2} \t^{\e/2} A_\e^{-1}
u_R.
\eea
A convenient choice for the normalization point NP is
${k}_i= \o_i =0$ and $\t =\mu^2$, where $\mu$ is an arbitrary momentum
scale. To one loop, we find 
\bea
\G_{11} (w, {\bf k}) &=& i\o +{e_0 \over 2}k_\pe^2 (k^2+\t)+D_1, \nonumber \\
\G_{13} (w, {\bf k} ) &=& {e_0 \over 2}k_\pe^2 g + D_2,
\eea
where $D_1$($D_2$) corresponds to the algebraic expression of 
the left(right) diagram in Figure \ref{diagdiv}.
A calculation in dimensional regularisation \cite{amit,GF} yields 

\bea
D_1&=& {5 g e_0\over 128 \pi^2} k_\bot^2 {\t^{1-\e/2} 
\over \e}, \nonumber \\
D_2&=& 
-{5 g^2 e_0\over 128 \pi^2} k_\bot^2 {\t^{-\e/2}  \over \e}.
\eea
After setting $A_\e = 5/64 \pi^2$, one obtains
\bea
\partial_{k_\bot^2} \G_{11} \vert_{NP} &=&
{e_0 \over 2} \t +  {5 g e_0\over 128 \pi^2} {\t^{1-\e/2}
 \over \e}= {e_0 \over 2} \t \bigg[1+ {5 \t^{-\e/2}
\over 64 \pi^2} {g \over \e}
\bigg]= {e_0 \over 2} \t \bigg[ 1+{u \over \e} \bigg], \nonumber \\
\partial_{k_\bot^2} \G_{13} \vert_{NP} &=&
{e_0 \over 2} A_\e^{-1} \t^{\e/2} u \bigg[1- {u \over \e}\bigg],
\eea
which entails
\bea
Z_\t= 1+{u \over \e} + O(u^2), \nonumber \\
Z_u= 1-{u \over \e}+ O(u^2).
\eea
The renormalization group equation obtained
after requiring invariance of the bare irreducible vertex functions
upon changes on the normalization point reads
\be
\Big[ \m \partial_\m +\b \partial_{u_R} + \z \partial_{\t_R}
\Big] \G_{\tilde n n}^R =0,
\ee
where the renormalization group functions are defined in the usual way:
$\b(u_R) \equiv \m \partial_\m u_R$, 
and $\z(u_R) \equiv \m \partial_\m (\ln \t_R)$.
A straightforward calculation then leads to 
\bea
\b (u_R) &=&-\e u_R + u_R^2 + O(u_R^3), \nonumber \\
\z (u_R) &=& 2-u_R +O(u_R^2),
\eea
from which one can determine the location and stability of the fixed
points. 
To this order, apart from the trivial mean-field result $u_R^*=0$, 
a nontrivial fixed point $u^*_R=\e$ emerges.
 This fixed point controls the
critical behavior of the theory below four dimensions. 

\begin{figure}
\begin{center} \begin{picture}(120,100)(0,0)
\Line(0,50)(30,50)
\Photon(30,50)(60,50){1}{8.5}
\Line(60,50)(90,50)
\Photon(90,50)(120,50){1}{8.5}
\CArc(60,50)(30,0,360)
\Vertex(30,50){2}
\Vertex(90,50){2}
\end{picture}
\end{center}
\label{dosloops}
\caption{Two loop contribution to $\G_{11}$.}
\end{figure}
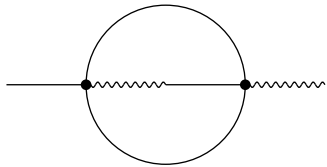

Now we proceed with the calculation of the associated critical exponents.
We first note that, as indicated above,
 no renormalization of the fields 
$\tilde \phi$, $\phi$ has been required. Therefore,
in particular, the anomalous dimension of $\phi$ vanishes, {\em i.e.\/}, 
$\eta =0$.
Concerning the exponent $\nu$, which controls the divergence of
the correlation length with temperature, we simply have 
$\nu= \zeta(u_R^*)^{-1}=1/2+\e /4 + O(\e^2)$.
This is to be compared with $\n =1/2$, the value obtained by Janssen  and 
Schmittmann for $E=\infty$ \cite{janssen}. 
This result demonstrates that
the continuous version Eq. (\ref{new})of the DLG with
 $E=\infty$ is not mean-field like but characterizes
a universality class other than the 
one in \cite{janssen}.
Since there are no dangerous
 irrelevant operators in Eq. (\ref{funcional}) 
 standard scaling laws apply (contrary to the case in \cite{janssen}).
Therefore the exponents are related to each other and
estimating $\eta$ and $\nu$ is sufficient to 
deduce all the other exponents.
 For instance, the order parameter exponent $\b$
can be written as
$\b ={\nu \over 2} (d-2+\eta)$ \cite{amit}, 
and we have $\b =1/ 2 +O(\e^2)$.

The previous results concern the one loop approximation.
The two loop calculation  
presents an important
new feature, namely, that the scaling becomes anisotropic.
The detailed two loop calculation 
involves the evaluation of many non trivial integrals and will be presented elsewhere. 
We do not present it here, but let us stress that
a simple glance to some of the contributing integrals permits us 
to extract far reaching consequences.
In fact,
Figure 3 reveals that, contrary to what happens
to one loop order, the graph contributing
to $\G_{11}$ to two loops depends on external frequencies and momenta.
Therefore, $e_0$ and the fields need to be renormalized to heal
these new divergences. Moreover, this
diagram depends on the parallel and perpendicular components of the 
external momenta in an asymmetric way; consequently, anisotropic 
exponents emerge. On the other hand, after including two loop
diagrams,  
corrections to the $\b=1/2$ mean field value may appear. 
Finally, we remark that to two loops usual
scaling laws may no longer be valid; in particular, the exponent
$\nu$ will split up in $\nu_{\pe}$ and $\nu_{\pa}$.

 Summing up, we have performed the renormalization of the 
field theory in \cite{pre} for the DLG under 
 an infinitely large driving field. The renormalization procedure
 yields results essentially different from those for a 
finite field. In particular, corrections to mean field
are observed explicitly in the one loop approximation for 
the exponent $\nu$. Anisotropic exponents and a non-mean-field exponent 
$\beta$ are predicted from simple
arguments based on the analysis up to two loop diagrams. This 
calls for extensive computational 
simulations to observe numerically the difference
between finite and infinite driving  cases.
{\bf Acknowledgements} It is a pleasure to acknowledge J. Marro,
J. L. Lebowitz and P.L. Garrido for useful discussions, and E. Hern\'andez-Garc{\'\i}a 
for pointing out reference \cite{nematic} to us.
 This
work has been partially supported by the 
European Network Contract ERBFMRXCT980183
 and by the Ministerio de Educaci\'on
under project DGESEIC, PB97-0842.

\end{document}